\documentclass[a4paper]{jpconf}
\usepackage{graphicx}
\usepackage{upgreek}

\begin{document}
\title{A fast and parametric digitization for triple-GEM detectors}

\author{R. Farinelli$^{1,2}$,
M. Alexeev$^3$,
A. Amoroso$^3$,
S. Bagnasco$^3$,
R. Baldini Ferrioli$^{4,5}$,
I. Balossino$^{1,5}$,
M. Bertani$^4$,
D. Bettoni$^1$,
A. Bortone$^{3,6}$,
F. Bianchi$^{3,6}$,
A. Calcaterra$^4$,
S. Cerioni$^4$,
J. Chai$^3$,
W. Cheng$^3$,
S. Chiozzi$^1$,
G. Cibinetto$^1$,
F. Cossio$^3$,
A. Cotta Ramusino$^1$,
G. Cotto$^{3,6}$,
M. Da Rocha Rolo$^3$,
F. De Mori$^{3,6}$,
M. Destefanis$^{3,6}$,
F. Evangelisti$^1$,
L. Fava$^3$,
G. Felici$^4$,
L. Gaido$^3$,
I. Garzia$^{1,2}$,
M. Gatta$^4$,
G. Giraudo$^3$,
S. Gramigna$^{1,2}$,
M. Greco$^{3,6}$,
L. Lavezzi$^{3,5}$,
S. Lusso$^3$,
H. Li$^3$,
M. Maggiora$^{3,6}$,
R. Malaguti$^1$,
A. Mangoni$^7$,
S. Marcello$^{3,6}$,
M. Melchiorri$^1$,
G. Mezzadri$^1$,
M. Mignone$^3$,
S. Pacetti$^7$,
P. Patteri$^4$,
B. Passalacqua$^{3,6}$,
A. Rivetti$^3$,
M. Savri\'e$^{1,2}$,
S. Sosio$^3$,
S. Spataro$^{3,6}$,
E. Tskhadadze$^8$,
L. Yan$^3$,
R.J. Wheadon$^3$}

\address{
$^1$ INFN Sezione di Ferrara, I-44122, Ferrara, Italy\\
$^2$ University of Ferrara, I-44122, Ferrara, Italy\\
$^3$ INFN Sezione di Torino, I-10125 Turin, Italy\\
$^4$ INFN Laboratori Nazionali di Frascati, I-00044, Frascati, Italy\\
$^5$ Institute of High Energy Physics, Beijing 100049, People's Republic of China\\
$^6$ University of Turin, I-10125 Turin, Italy\\
$^7$ INFN and University of Perugia, I-06100, Perugia, Italy\\
$^8$ Technological Institute of Georgia, Tbilisi, Georgia}

\ead{rfarinelli@fe.infn.it}

\begin{abstract}
Triple-GEM detectors are a well known technology in high energy physics. In order to have a complete understanding of their behavior, in parallel with on beam testing, a Monte Carlo code has to be developed to simulate their response to the passage of particles. The software must take into account all the physical processes involved from the primary ionization up to the signal formation, e.g. the avalanche multiplication and the effect of the diffusion on the electrons. In the case of gas detectors, existing software such as Garfield already perform a very detailed simulation but are CPU time consuming. A description of a reliable but faster simulation is presented here: it uses a parametric description of the variables of interest obtained by suitable preliminary Garfield simulations and tuned to the test beam data. It can reproduce the real values of the charge measured by the strip, needed to reconstruct the position with the Charge Centroid method. In addition, particular attention was put to the simulation of the timing information, which permits to apply also the micro-Time Projection Chamber position reconstruction, for the first time on a triple-GEM. A comparison between simulation and experimental values of some sentinel variables in different conditions of magnetic field, high voltage settings and incident angle will be shown.
\end{abstract}

\section{Introduction}
A software environment able to reproduce the triple-GEM performance is needed to extend the knowledge of the detector beyond the configurations studied in the test beam. The triple-GEM technology has been invented by F. Sauli in 1997 \cite{Sauli}. The technology is a gas detector that amplifies the charge of the primary ionization produced by charged particles interacting with it. The amplification is given by three foils of GEM. Each GEM is built up by a thin coppered kapton foil with thousands of holes of 50 $\upmu$m. Hundreds of Volts are applied between the two faces of the foil and an intense electric field is generated in the holes. An electron entering the hole is accelerated and ionizing the gas generates an electron avalanche. Three GEM foils allow to reach a total gain of 10$^4$-10$^5$. An electric field outside the GEM foils make the electron drift from the cathode to the anode.

The simulation of the detector and its signal generation is divided in several steps: gas ionization, electron diffusion in gas volume, GEM amplification and signal induction on the anode. At present, Garfield++ \cite{Garfield} is a complete software that reproduces the gas detector signal generation with high accuracy. Unfortunately Garfield++ is very time consuming and there is the need to develop a faster digitization of the detector. A new software with a smaller time consumption, named GTS (Garfield-based Triplegem Simulator), is proposed in this article. 

The basic approach shown in this work shares the idea in Ref. \cite{Bonivento}. The method has been extended in a wider range of configurations to take into account the effects of the detector gain, incident angle of the tracks and the magnetic field. The simulated events have been reconstructed with the same technique of the experimental data: the signal charge and time information are extracted and used to measure the incident position with the charge centroid (CC) and $\upmu$TPC algorithms \cite{RF}. 

The simulation of the triple-GEM is studied through independent processes that can be described separately:
\begin{itemize}
\item ionization: interaction of a charged particle with the gas medium and generation of primary and secondary electrons;
\item electron drift properties: electron motion depending on the electron and magnetic fields and the gas mixture;
\item GEM properties: gain and GEM transparency;
\item induction of the signal: current generated by electron motion and its readout by the electronics.
\end{itemize}

The description of the electric field map inside the triple-GEM detector has been performed with Ansys software \cite{Ansys}.

\section{Primary ionization simulation}
\begin{figure}[tp]
  \centering
  \begin{tabular}{ccc}
    \includegraphics[width=0.28\textwidth]{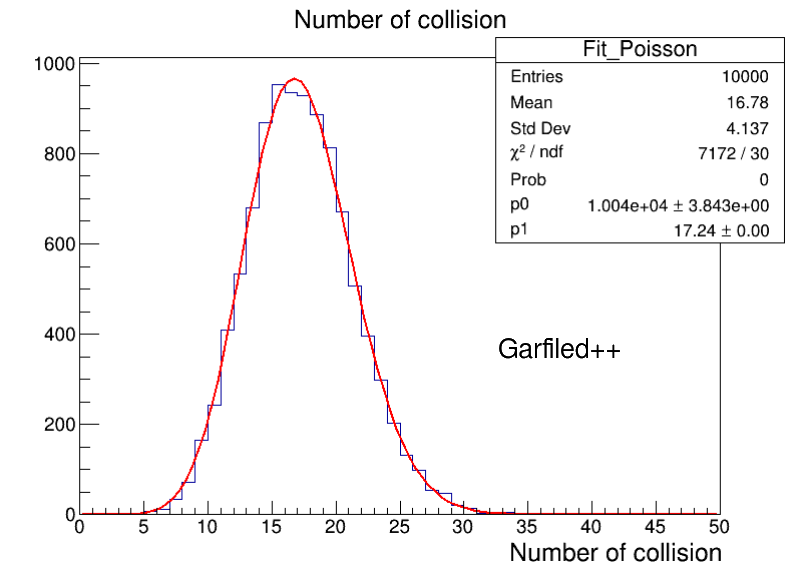}&
    \includegraphics[width=0.28\textwidth]{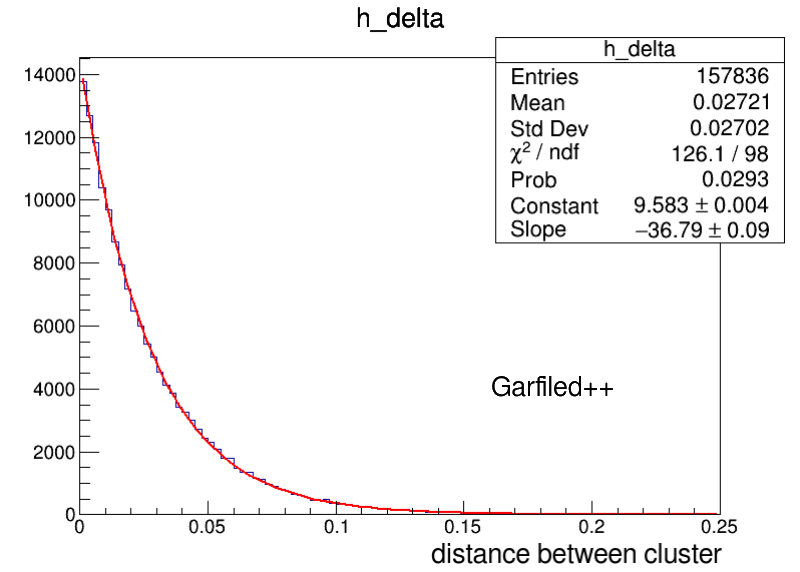}&
    \includegraphics[width=0.28\textwidth]{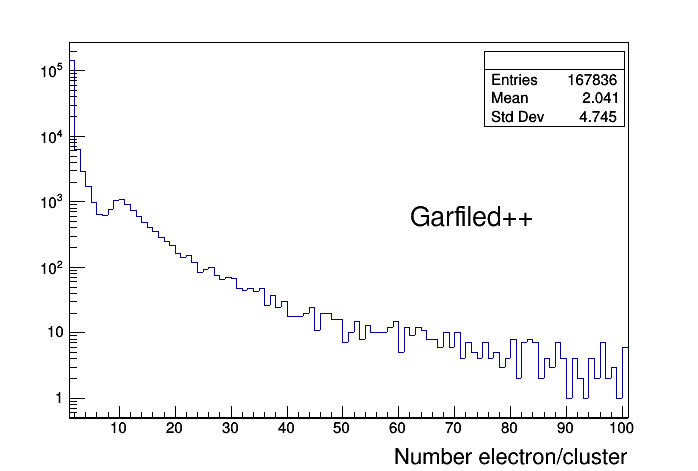}\\
    \includegraphics[width=0.28\textwidth]{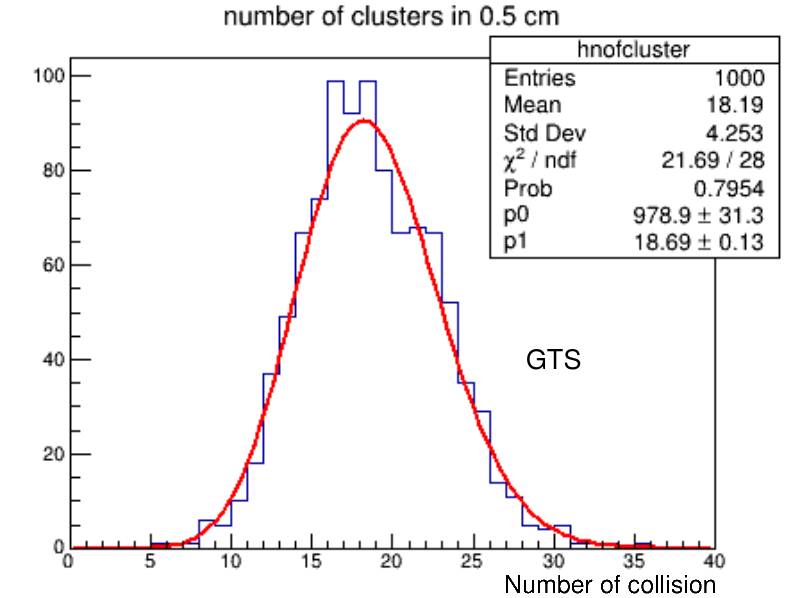}&
    \includegraphics[width=0.28\textwidth]{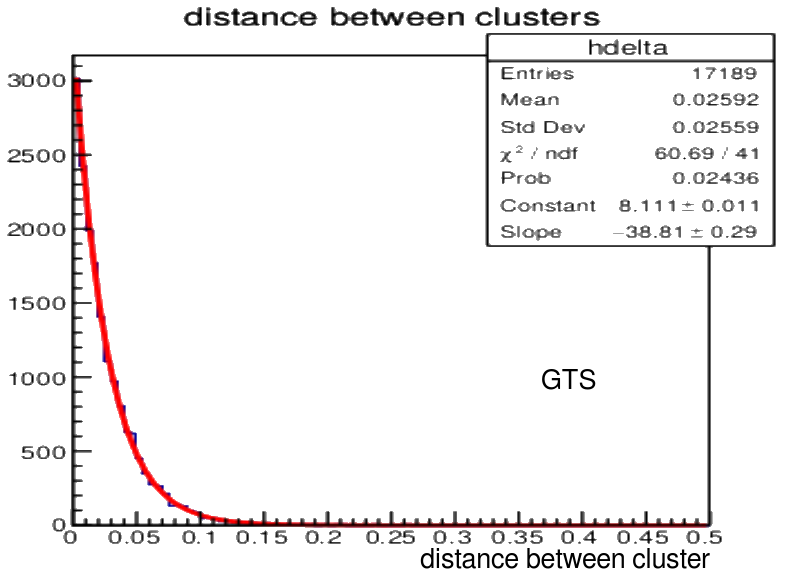}&
    \includegraphics[width=0.28\textwidth]{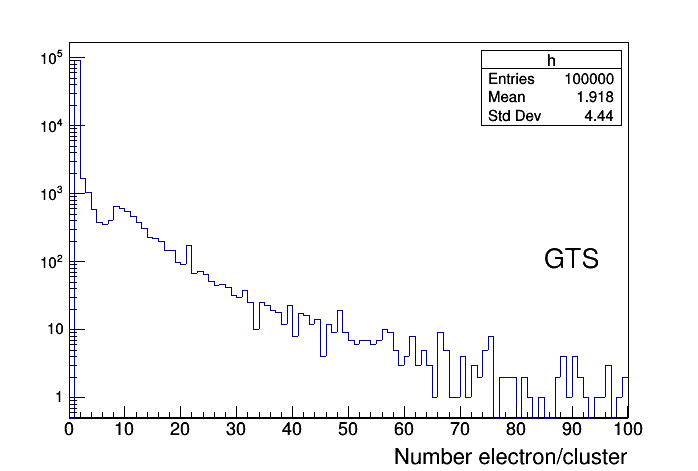}\\
  \end{tabular}
  \caption{Distributions of the processes on the ionization from Garfield++ (top) and GTS (bottom). The first column shows the number of cluster, the second column the distance between each cluster along the ionizing particle path and the third column the number of secondary electrons in the electron cluster.}
\label{fig:digi_ioni}
\end{figure}

Heed \cite{Heed} is a Garfield++ component that generates ionization patterns through a model based on photo-absorption and ionization. Heed has been used to parametrize the number of primary electrons and their position. If the electron has sufficient energy then secondary electrons are generated. An agglomerate of electrons is named {\it electron cluster}. Relativistic muons have been shot with Garfield++ on a volume of Ar+10$\%$iC$_4$H$_{10}$ to measure and parametrize the number of electron clusters, the number of electron in a cluster and the relative distance between two consecutive cluster. Fits on those distributions have been used to import them in the digitization of the ionization. Fig. \ref{fig:digi_ioni} shows a comparison between Garfield++ and GTS.

\section{Electron diffusion}
\begin{figure}[tp]
  \centering
  \begin{tabular}{cc}
    \includegraphics[width=0.28\textwidth]{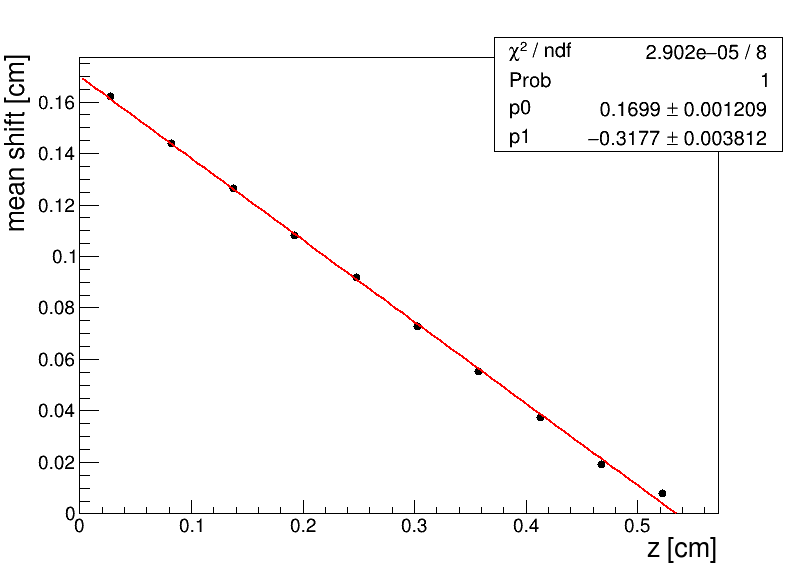}&
    \includegraphics[width=0.28\textwidth]{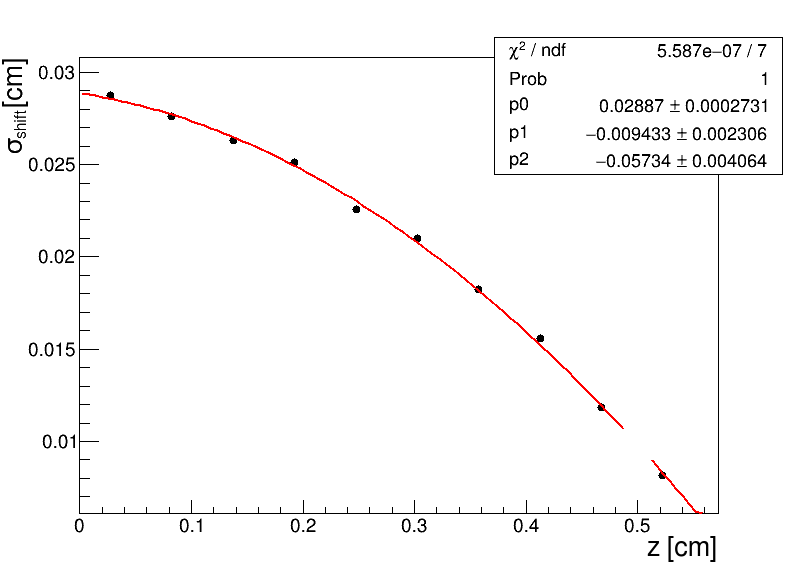}\\
    \includegraphics[width=0.28\textwidth]{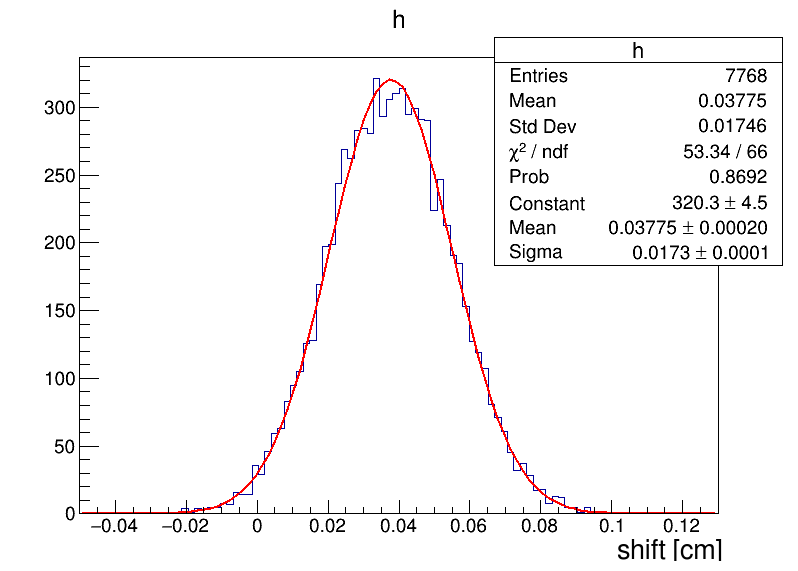}&
    \includegraphics[width=0.28\textwidth]{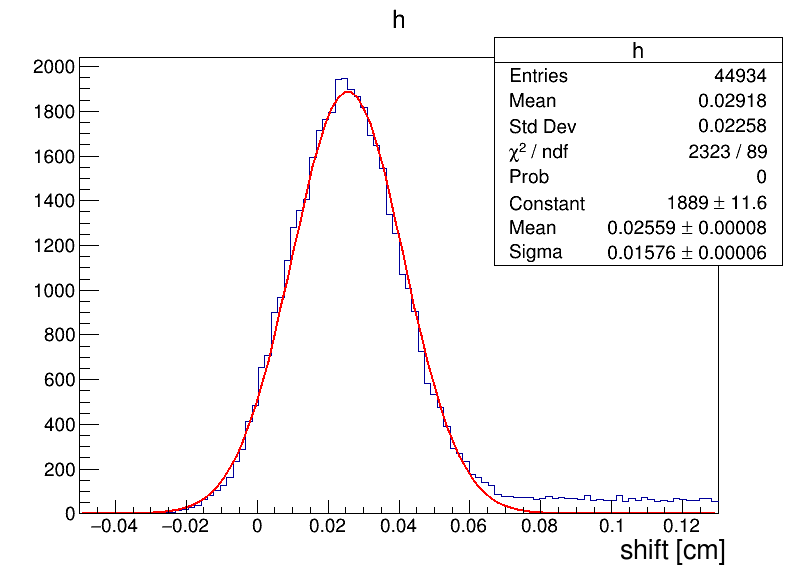}\\
  \end{tabular}
  \caption{Electron drift properties for all the gaps in a triple-GEM. The first row shows the mean shift (left) and spread (right) in the drift gap as a function of the distance $z$ from the cathode. In the second row the electron distributions in the transfer (left) and in the induction (right) gaps with the Gaussian fit are shown.}
\label{fig:digi_drift}
\end{figure}

The electron diffusion in the gas medium is an important parameter for a triple-GEM. Four different regions are present: a {\it drift gap} between cathode and the first GEM, two {\it transfer gaps} between the first two GEMs and between the last two GEMs, an {\it induction gap} between the third GEM and the anode. The electrons generated in the drift gap are amplified three times and their average path varies from 6 to 11 mm. This spreads their signal on a large region and Garfield++ has been used to parametrize the diffusion in space and time. The parametrization studies the gaps separately with simulations of the electron drift. A Gaussian fit has been used to measure mean shift and spread, as shown in Fig. \ref{fig:digi_drift}. A similar study has been done for the temporal diffusion.
\section{GEM properties}
The electric field inside and outside a GEM defines the intrinsic gain of the GEM and their transparency. Higher values of electric field inside the GEM foil increases the intrinsic gain while the electric field outside the GEM hole determines the fraction of the electrons that exit from it: the transparency. This quantities are related to the electrical configuration and to the electric field lines: if a larger number of lines falls on the GEM foil instead of entering the hole, then the transparency of the detector is low. The product of the intrinsic gain times the transparency returns the effective gain, as shown in Eq. \ref{eq:trasparency_gain}. A simulation of the electron drift and their amplification has been performed to compute the number of electrons entering the hole, $i.e.$ {\it collection efficiency}, and the number of electrons exiting,  {\it extraction efficiency}. 
\begin{equation}		
\begin{tabular}{ccc}
$G_{\mathrm{eff}}=G_{\mathrm{intr}} \times T$ & and & T=$\varepsilon_{\mathrm{coll}} \, \varepsilon_{\mathrm{extr}}$\\
\end{tabular}
\label{eq:trasparency_gain}
\end{equation}
The gain distribution of each GEM has been fitted with a Polya function and parametrized to be used in GTS.

\section{Signal induction}
The number of electrons generated in the drift gap is multiplied by a gain factor from the Polya distribution and the position in space and time of the primary electron is spread and shifted using the results of the previous sections. The anode plane is segmented in strips. The signal induction has been studied in two different ways: the technique of the Ramo theorem \cite{Ramo} and an approximation that can guarantee the same results. The first method simulates the drift of the electron up to the third GEM then it calculates the current induced on the strips as:
\begin{equation}		
I_k=-q \mathbf{v} \cdot \mathbf{E}_k^{\mathrm{weight}}
\label{eq:ramo_current}
\end{equation}
where $k$ is the strip index, $q$ and $\textbf{v}$ the electron charge and velocity, $\mathbf{E}_k^{\mathrm{weight}}$ the weighting field evaluated as the electric field generated by the k-strip if 1 V is applied on it and 0 V on the others. The evaluation of the weighting field has been done with Ansys. The readout plays an important role in the simulation: a RC integration constant of 50 ns simulates the APV25 electronics \cite{APV} used in the experimental setup and the current is sampled every 25 ns and it is digitized in 1800 ADC. Thanks to the large number of electrons reaching each strip, another method has been developed in order to reduce the time consumption of the simulation. The induction process is neglected and the charge measured by the strips is given by the number of electrons reaching the anode as a function of their arrival time. This approach strongly improves the time performance of the simulation.
A fluctuation of the current recorded by the strip is introduced to simulate the electronic noise.
\section{Tuning and comparison of the simulation to the experimental data}
\begin{figure}[tp]
  \centering
  \begin{tabular}{cc}
    \includegraphics[width=0.28\textwidth]{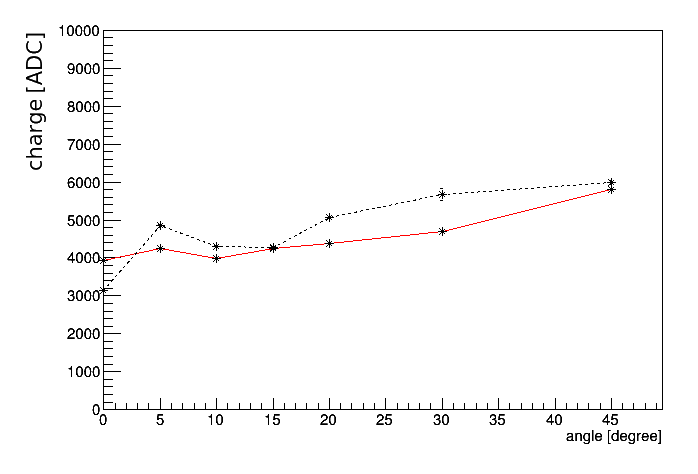}&
    \includegraphics[width=0.28\textwidth]{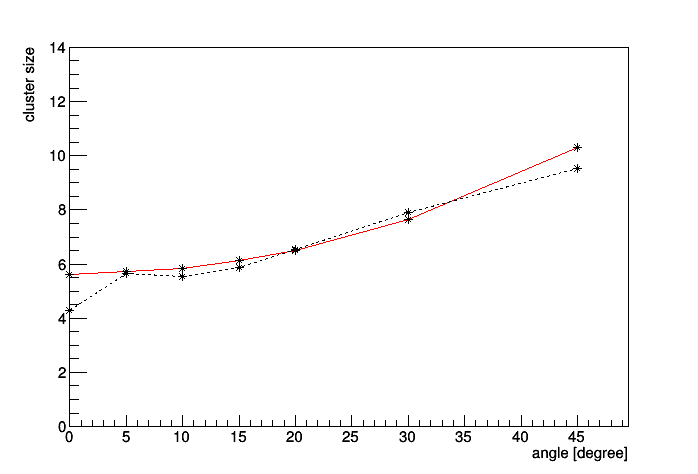}\\
    \includegraphics[width=0.28\textwidth]{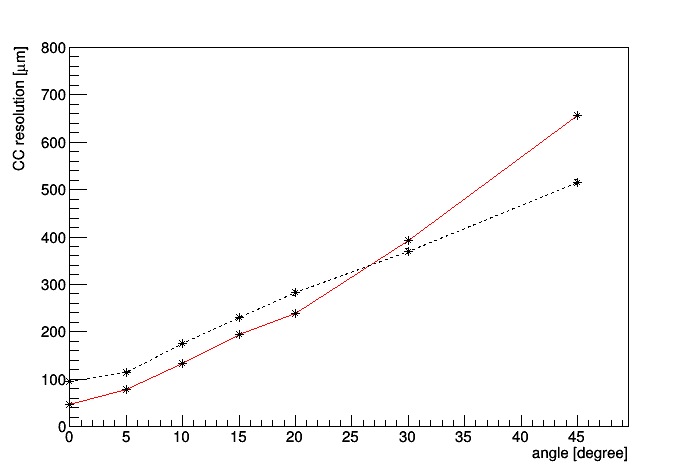}&
    \includegraphics[width=0.28\textwidth]{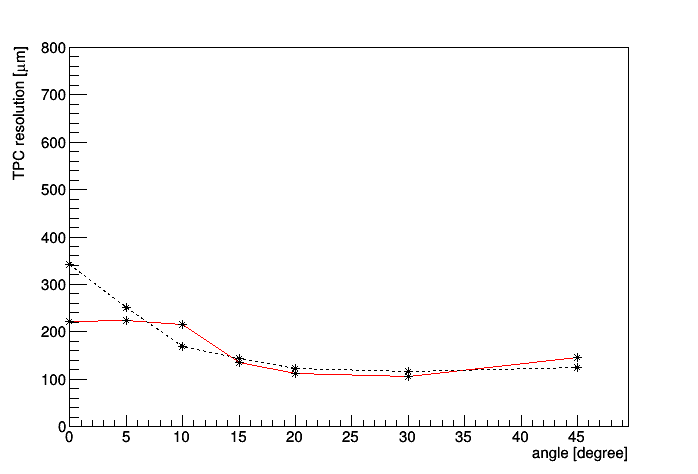}\\
  \end{tabular}
  \caption{Experimental data (black line) and simulations (red line) are compared as a function of the incident angle. On the top left the charge of the signal collected on the strip is shown, on the top right the number of fired strips. In the bottom line is shown the spatial resolution of the detector with the two reconstruction algorithms: CC (left) and $\upmu$TPC (right) is shown.}
\label{fig:comparison}
\end{figure}

The GTS environment has been used to simulate the response of a triple-GEM detector in several conditions, in order to compare the performance measured in a test beam. The settings used in the simulation are the same of the experimental setup. A high voltage scan has been used to test the behavior of the GEM gain and the transparency while an incident angle was used to test the reconstruction algorithms of the CC and $\upmu$TPC. Magnetic field has been studied too. In order to achieve an agreement between data and simulation within 30$\%$ two variables needed a {\it tuning}. The gain simulated by Garfield++ needs a factor 4.5 to be fixed. A factor 2 to fix this problem is a known problem \cite{GarfGain} in the MPGD comunity, the larger value used is due to the fast induction technique. The electron diffusion is the other parameter to be tuned in order to have a good agreement of the $\upmu$TPC at small incident angle. A factor 2 has been introduced to complete the tuning. The comparison of the data and simulation is shown in Fig. \ref{fig:comparison}.
\section{Conclusion}
A fast triple-GEM simulator has been developed to reproduce the performance of the detector. This software uses the parametrization of variables from Ansys and Garfield++. It can improve significantly the time consumption of a triple-GEM simulation to a fraction of second. In the parametrization the different processes have been treated independently. The variables of interest used in the simulation have been tuned using data collected in test beam. The agreement of the simulation and data is better than 30$\%$ and the method can be exploited in several applications such as the simulation of a triple-GEM detector in large environment such as the BESIII experiment.

\end{document}